\documentclass[12pt]{article}
\usepackage{amsmath,amssymb,physics,booktabs}
\usepackage{graphicx}
\usepackage{bm}
\usepackage{geometry}
\usepackage{hyperref}
\usepackage{authblk}
\usepackage{enumitem}
\usepackage{physics}
\usepackage{tikz}
\usepackage{hyperref}
\usepackage{tikz-3dplot}
\usepackage{caption}

\title{Contextuality, Incompatibility,  and Intra-System Entanglement of Mental Markers}

\author[1]{Andrei Khrennikov \thanks{andrei.khrennikov@lnu.se}}
\author[2,3]{Felix Benninger}
\author[3,3]{Oded Shor}

\affil[1]{Center for Mathematical Modeling in Physics and Cognitive Sciences, Linnaeus University, V\"axj\"o, SE-351 95, Sweden}
\affil[2]{Felsenstein Medical Research Center, Petach Tikva, Israel}
\affil[3]{Sackler Faculty of Medicine, Tel Aviv University, Tel Aviv, Israel}

\date{}

\begin{document}
\maketitle

\begin{abstract}
Over the past two decades, quantum-like modeling (QLM) has emerged as a powerful framework for describing non-classical features of cognition and decision-making. Rather than assuming physical quantum processes in the brain, QLM employs the Hilbert space formalism to model contextuality, incompatibility of mental observables, and entanglement-like correlations. In this paper, we develop a quantum-informational model of mental markers within the broader $I$-field (information field) approach. We propose that, under conditions of information overload and limited cognitive resources, individuals primarily respond not to detailed semantic content but to compact content labels—mental markers—carrying cognitive and affective components. We formalize mental markers as structured quantum-like states and analyze the nonclassical correlations between their cognitive and affective components using the {\it Contextuality--Incompatibility--Entanglement triad.} Special attention is given to intra-system entanglement between rational (cognitive) evaluation and emotional (affective) coloring, accounting for context-dependent judgments, order effects, and affect-driven decision shifts. Illustrative examples with psychological interpretation and experimental perspectives are provided. An Appendix briefly discusses neurobiological analogues of information overload in neural networks, highlighting structural parallels with the proposed marker-based framework; coupling to the origin and diagnostics of neurological diseases is analyzed.  The paper contributes to QLM by distinguishing inter-system and intra-system entanglement and by demonstrating that cognitive--affective entanglement constitutes a fundamental structural feature of mental markers in socially mediated information environments.
\end{abstract}

{\bf keywords:} quantum-like cognition and decision making; mental markers; entanglement of cognitive and affective markers;
 inter- and intra-system entanglement; contextuality–incompatibility–entanglement triad; 
neurological diseases; information overload in neural networks

\section{Introduction}

The rapid development of quantum information theory—often referred to as {\it the second quantum revolution}—has led  not only to technological achievements (as quantum computing and cryptography), but also to profound foundational insights regarding the role of quantum theory in contemporary science. Quantum information theory highlights the probabilistic and information-theoretical aspects of quantum mechanics, extending its significance far beyond the boundaries of traditional physics.  

One of the most significant outcomes of this paradigm shift is the emergence of {\it quantum-like modeling (QLM)} —the application of quantum theoretical formalism and methodology to domains outside physics, including cognition, psychology, decision-making, 
sociopolitical sciences, economics and finance, game theory, and artificial intelligence,  biology including genetics and epigenetics, and evolution theory (see, see for example, monographs \cite{QL0,UB_KHR,Busemeyer,QL3,Bagarello,Bagarello1,Open_QLM}, reviews
\cite{Pothos1,BioBas}, and some articles which are closer related to this paper \cite{Aerts1}-\cite{Aerts3},
\cite{Pothos,Busemeyer4,AT1,AT2,AT3,Bruza,Bagarello2,Busemeyer5,Gunji3,Gunji4,Wang1}). 
This conceptual expansion from quantum physics to QLM has been aptly termed {\it the quantum-like revolution} \cite{Zeit}.

This is the good place to emphasize that the cognitive counterpart of QLM, {\it quantum-like cognition and decision making} (QCDM) 
should be sharply distinguished from the quantum brain project (in the spirit of Umezawa, Vitiello, Hameroff, and Penrose). The latter is based on the reductionist paradigm - the reduction of mental phenomena to genuine quantum physical processes in the brain. QLM is based on the paradigm that information processing in macroscopic systems, say humans, ecological, social, or political systems can violate the laws of classical probability and information. And the quantum formalism can be used to resolve corresponding inconsistencies, e.g. the probability fallacies, paradoxes of decision theory, irrational behavior of agents acting in various contexts.\footnote{We make the following terminological remark. Some authors, e.g., Aerts, Busemeyer, Bruza, use the terminology 
``quantum cognition and decision making'' in the QLM framework. Both terminologies are well established in QLM. 
 The abbreviation QCDM matches both terminologies.}      

The success of QCDM is rooted in its ability to model \textit{contextuality} and \textit{incompatibility}. Traditional rational choice models often fail to account for cognitive biases, such as the conjunction fallacy or order effects, which emerge because mental observables often do not commute \cite{Busemeyer,Pothos1,Busemeyer4,Wang1}.\footnote{Further QCDM-studies demonstrated that by attempting to combine a few psychological effects, e.g., the order and response replicability effect \cite{PLOS}, we should use more advanced  mathematical formalism - theory of open quantum systems and quantum instruments \cite{OK20,OK21,OK23,Fuyama}.} In QLM, an agent's mental state is a superposition of potential outcomes, and the transition to action is modeled as a ``state collapse'' driven by the evaluation of \textit{mental markers}.   

Nowadays, people interact intensively with mass media, social networks, and a wide variety of online platforms. In modeling these interactions, it is useful to employ the notion of an information field, representing the information absorbed by individuals. In certain situations---such as information overload \cite{Pothos1} or generally limited time and cognitive resources---people are unable to analyze messages in detail. Instead, they often respond to \textit{excitations of the information field}, which carry portions of social (psychic, mental) energy along with \textit{mental markers} that compactly encode the content of messages. One can think of these as \textit{quanta of the information field}, each associated with a portion of social energy and a content label (mental marker).  

Within this framework, it is natural to represent the information field \(F\), or ``\(I\)-field,'' using \textit{quantum field theory}. In  \cite{SLTarXiv} the quanta of the \(I\)-field were termed \textit{``infons''}, analogous to photons in the electromagnetic field. Mathematically, a quantum field is an operator-valued generalized function (distribution). It can be represented as a combination of \textit{creation and annihilation operators}, which respectively generate and remove quanta of the field. For the \(I\)-field \(F\), these quanta correspond to portions of social energy coupled with mental markers.  

The central idea of the \(I\)-field approach is to reduce the enormous variability of information forms to a discrete set of labels, \((E, m)\), where the symbols  $E$ and $m$  denote social (psychic) energy and mental marker.   The fundamental psycho-informational insight is that, in interacting with \(F\), humans primarily ``perceive'' these labels rather than the full content. We emphasize that this applies to situations where individuals do not carefully analyze content or reason logically---they register only the content labels and the associated level of excitation.  

This perspective underlies the \textit{Social Laser Theory (SLT)}, developed in a series of works \cite{laser1}-\cite{laser8}, \cite{laser12,laser13}. SLT focuses on modeling flows of social energy and the generation of a social analog of laser-like regimes, stimulating coherent social actions. While content markers were only briefly discussed in earlier SLT study \cite{SLTarXiv}, they play a crucial role in decision-making processes. The present article focuses specifically on the \textit{quantum informational structure of mental markers}. Our goal is not to model the induction of coherent social actions, but to analyze the structural organization of content-carrying markers.  

In our model, each mental marker consists of \textit{cognitive and affective components}, and we study correlations between these components.\footnote{Unlike previous works, we do not explicitly invoke the notion of social energy, which remains vague and experimentally untested. The notion of an affective marker is closer to established concepts in cognitive science and decision-making and is therefore more amenable to systematic study.} Our aim is to demonstrate how the nonclassical structure of these correlations can be modeled within a quantum formalism, based on the \textit{Contextuality--Incompatibility--Entanglement (CIE) triad}. We illustrate the entanglement between cognitive and affective markers with numerous examples, some of which could, in principle, be developed into experimental tests. Each example is accompanied by a corresponding psychological interpretation.  

In addition to its cognitive and social applications, the present QLM framework has potential medical relevance (see also \cite{Shor}). In Appendix B, we analyze several neurological and neuropsychiatric conditions---including migraine, epilepsy, autism spectrum disorder, ADHD, and Alzheimer’s disease---as cases of pathological information overload and impaired filtering. We argue that failures of neural habituation, excitatory--inhibitory imbalance, and hypersynchronization can be interpreted as breakdowns of the brain’s capacity to compress rich input into stable marker-like representations. From this perspective, the theory of mental markers and social lasing provides a unified informational language for describing threshold phenomena, phase transitions, and loss of contextual differentiation in both social and neural systems. Such a cross-level interpretation suggests possible conceptual bridges between quantum-like cognitive modeling and network-based approaches in clinical neuroscience.  

This paper highlights two forms of entanglement in quantum physics and their expression in QLM. In quantum physics, one distinguishes \textit{inter-system and intra-system entanglement}. Inter-system entanglement occurs between degrees of freedom of two subsystems of a composite system, originally considered by Schr\"odinger in connection with the EPR paradox, and later explored by Bell \cite{Bell} with emphasis on quantum nonlocality. Another type of entanglement, \textit{intra-system entanglement}, occurs between degrees of freedom within a single system and has been extensively studied in quantum physics, especially in optics \cite{IE1,IE2,IE3,IE4,IE5}.  

Such a classification of entanglement has not yet been systematically applied in QLM (with the exception of \cite{Yamada}, which briefly mentions a double structure of mental entanglement). Most studies in QLM have focused exclusively on intra-system entanglement.  

In this work, we pay particular attention to \textit{intra-system entanglement between cognitive and affective components of mental markers}. This entanglement implies that the rational evaluation of a social situation is \textit{contextually linked to its emotional coloring} (cf. \cite{emotion}). In other words, cognitive judgments and affective responses cannot, in general, be represented as separable subsystems; rather, they form a unified quantum-like structure in which changes in one component alter the probabilistic structure of the other. By making this structure explicit, we aim to clarify the mathematical architecture of mental markers and to provide a systematic foundation for modeling context-sensitive decision dynamics in socially mediated information environments, with possible coupling to neuro-physiological processes in the brain.  

\section{Mental Markers: Cognitive versus Affective}
\label{CAM}

We introduce mental markers in the following way: 

\medskip

\textbf{Definition.} A \emph{mental marker} is a semantically charged element embedded in a communicative act---such as a word, symbol, headline, tone, or visual cue---that triggers automatic interpretative or emotional responses by activating culturally or psychologically encoded meaning patterns.  These markers guide interpretation, frame meaning, and facilitate associative processing by activating pre-existing mental schemas, ideological frameworks, or affective dispositions.
 the receiver.

\medskip

A {\it cognitive marker} is a feature, symbol, or informational element that plays a role in
structuring and guiding thinking, reasoning, knowledge acquisition, and interpretation. 
Cognitive markers are primarily associated with knowledge, beliefs,  perceptions, memory, mental models, decision-making, and problem-solving strategies. People operate with them in information processing, understanding, and mental representations. 
In the context of SLT or QCDM, cognitive markers may be linguistic cues, images, concepts, or framing devices that organize
mental representations and influence decision-making.

An {\it affective marker} is a cue or feature that encodes emotional valence (positive or 
negative) and/or arousal, shaping motivation, attention, and memory. They involve emotional 
responses, feelings, and moods like happiness, sadness, anger, fear, or anxiety, as well as 
subjective experiences -  how we feel about things.

In neuroscience, Damasio's somatic marker hypothesis \cite{Damasio} treats affective markers as emotional ``tags'' that bias
reasoning by associating feelings with certain options or stimuli. They are primarily
concerned with emotional evaluation and motivation, rather than purely propositional
reasoning.

{\bf Interconnectedness:} Cognitive and affective processes are not isolated; they influence each other. For example, a negative thought (cognitive) can trigger a feeling of sadness (affective), and conversely, a sad mood can influence how we interpret information (cognitive).

{\bf Impact on behavior:} Both cognitive and affective markers play a role in shaping our actions and reactions.

{\bf Differences:}
\begin{itemize}
\item Nature of the process:
Cognitive markers are primarily about thinking and reasoning, while affective markers are about feeling and emotion. 
\item Focus of study:
Cognitive psychology often focuses on understanding how information is processed, while affective psychology focuses on understanding emotions and their impact. 
\item Examples:
Cognitive markers include things like memory and problem-solving, while affective markers include things like joy and fear. 
\end{itemize}
In essence, cognitive markers are about ``thinking,'' and affective markers are about ``feeling.'' They are distinct yet intertwined aspects of human experience. 

 {\bf Examples in SLT terms:}
\begin{itemize} 
\item Cognitive marker: ``democratic reform'' — orients the listener toward a political
concept.
\item Affective marker: ``urgent fight for our freedom'' - overlays strong emotional
valence.
\end{itemize}

A QLM of emotional coloring of cognitive experiences was suggested in article \cite{emotion}; it matches well with Damasio's somatic marker hypothesis \cite{Damasio}.

In practice, many markers (e.g., ``fight for our freedom'') are both cognitive (encoding
a political action frame) and affective (inducing emotional arousal). Therefore it is convenient  to introduce a new term a {\it mental marker} and consider its cognitive and affective components, $m=(c,a).$ 

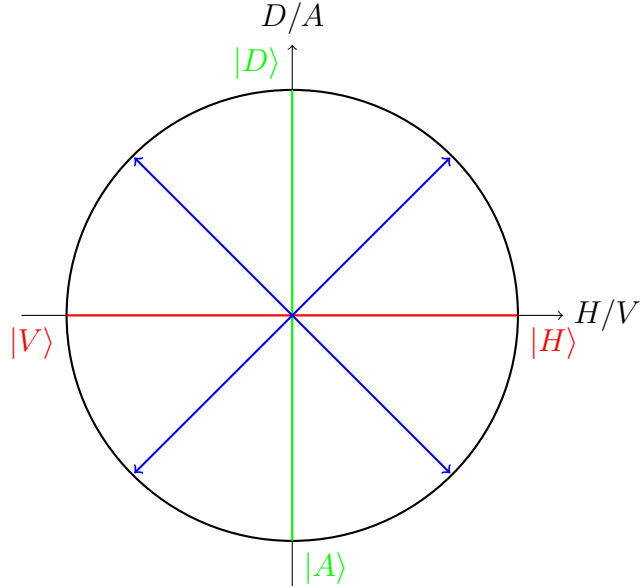
\begin{figure}[h!]
    \centering
    \begin{tikzpicture}[scale=3]

    \draw[thick] (0,0) circle (1);

    \draw[->] (-1.2,0) -- (1.2,0) node[right] {$H/V$};
    \draw[->] (0,-1.2) -- (0,1.2) node[above] {$D/A$};

    \draw[red, thick] (1,0) node[below right] {$|H\rangle$} -- (0,0);
    \draw[red, thick] (-1,0) node[below left] {$|V\rangle$} -- (0,0);
    \draw[green, thick] (0,1) node[above left] {$|D\rangle$} -- (0,0);
    \draw[green, thick] (0,-1) node[below right] {$|A\rangle$} -- (0,0);

    \draw[->, blue, thick] (0,0) -- (0.7,0.7);
    \draw[->, blue, thick] (0,0) -- (-0.7,0.7);
    \draw[->, blue, thick] (0,0) -- (0.7,-0.7);
    \draw[->, blue, thick] (0,0) -- (-0.7,-0.7);

    \end{tikzpicture}
    \caption{2D circle representing in-plane photon polarizations. Linear polarizations: horizontal ($\lvert H \rangle$) and vertical ($\lvert V \rangle$); diagonal polarizations: diagonal ($\lvert D \rangle$) and anti-diagonal ($\lvert A \rangle$). Blue arrows indicate superpositions from the center toward the diagonal directions.}
\label{figPolarization}
\end{figure}

\begin{figure}[h!]
    \centering
    \begin{tikzpicture}[scale=3]

    \draw[thick] (0,0) circle (1);

    \draw[->] (-1.2,0) -- (1.2,0) node[right] {Valence (Negative → Positive)};
    \draw[->] (0,-1.2) -- (0,1.2) node[above] {Arousal (Low → High)};

    \node at (0.7,0.7) {Excited / Joy};    
    \node at (-0.7,0.7) {Angry / Fear};    
    \node at (0.7,-0.7) {Calm / Content};  
    \node at (-0.7,-0.7) {Sad / Bored};    

    \draw[->, blue, thick] (0,0) -- (0.7,0.7);
    \draw[->, blue, thick] (0,0) -- (-0.7,0.7);
    \draw[->, blue, thick] (0,0) -- (0.7,-0.7);
    \draw[->, blue, thick] (0,0) -- (-0.7,-0.7);

    \end{tikzpicture}
    \caption{Circumplex model of emotions. Horizontal axis: valence (negative to positive); vertical axis: arousal (low to high). Blue arrows indicate positions of example emotions: high-arousal positive (excited/joy), high-arousal negative (angry/fear), low-arousal positive (calm/content), low-arousal negative (sad/bored).}
		\label{figEmotions}
\end{figure}
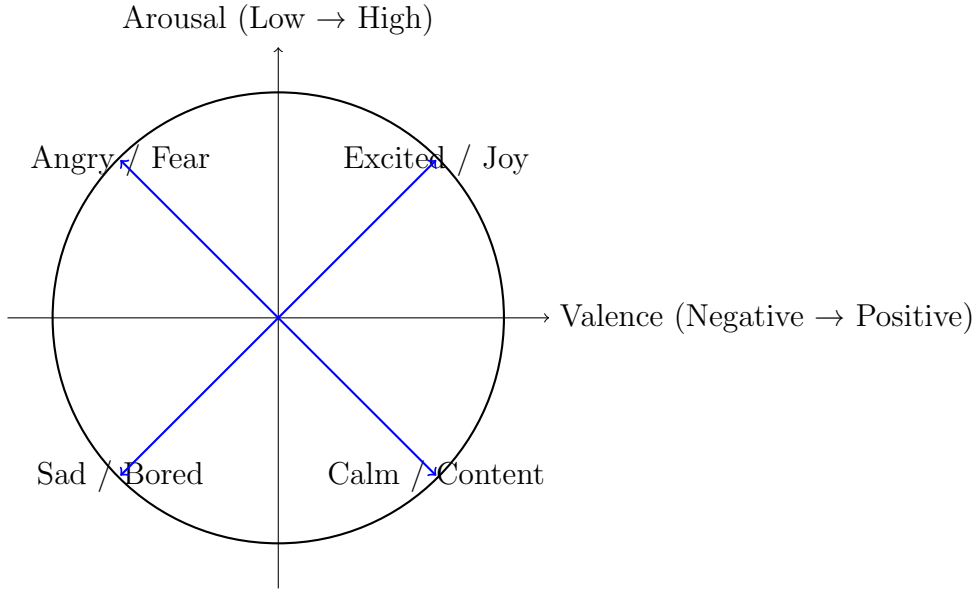

We note that the circle representation of polarization states of photons at Fig. \ref{figPolarization} matches the circle representation of emotion states at Fig. \ref{figEmotions}  as in articles \cite{Russell,Posner,Tseng, Jaeger,Wagner} (see 
also \cite{Tsarev}).

The state space for photon's polarization has complex dimension two and in SLT such space provides a rough picture of the emotioonal 
space, $\lambda= \pm$ - positive and negative emotions and they can be combined in superpositions. In more advanced models 
we can consider large values of the spin degree of freedom with multidimensional state spaces.      

\subsection{Cognitive-Affective Decomposition of Mental Markers}

Denote the sets of mental markers by the symbol $M.$ Here  $M=C\times A,$ where $C$ and $A$ are sets of cognitive (rational) and affective (emotional) markers. In the quantum model these sets are represented by complex Hilbert spaces ${\cal H}_M, {\cal H}_C, 
{\cal H}_A$ and ${\cal H}_M= {\cal H}_C \otimes {\cal H}_A,$ tensor product of Hilbert spaces for cognitive and affective markers.      
All mental markers are represented by pure states, for $m \in M, \; |m\rangle \in {\cal H}_M,$ and $\langle m|m\rangle =1.$

Let $m_1,m_2 \in M.$ Generally they are not orthogonal and their scalar product $\langle m_1|m_2\rangle$ gives the degree of correlation between them. Due to the tensor product decomposition, $\langle m_1|m_2\rangle= \langle c_1|c_2\rangle 
\langle a_1|a_2\rangle,$ that presents trade-off between cognitive and affective correlations.

\paragraph{Illustrative Examples of  Mental Marker Decomposition}

\paragraph{Example 1:``fight for our freedom''}  
We represent the mental marker as a pair which coordinates are cognitive and affective markers respectively 
\[
m_1 = (c_1,a_1), \qquad m_1 \in M=C \times A.
\]
\textbf{Cognitive marker ($c_1$):} encodes a \emph{political action frame}, namely the idea of collective struggle oriented toward achieving or preserving freedom.  
\[
c_1 = \text{``collective struggle for freedom''}.
\]
 \noindent \textbf{Affective marker ($a_1$):} overlays urgency, solidarity, pride, and anger, i.e. a high-arousal emotional charge.  
\[
a_1 = \text{``urgent solidarity / pride / anger''}.
\]
\noindent  \noindent \textbf{Representation in ${\cal H}_M$:}
\[
|m_1\rangle = |c_1\rangle \otimes |a_1\rangle 
= |\text{struggle for freedom}\rangle \otimes |\text{urgent solidarity/pride/anger}\rangle.
\]
 The affective marker component of this mental marker is essentially stronger than its cognitive component, so it can be treated as an affective marker. 

\paragraph{Example 2: ``democratic reform''}  

\[
m_2 = (c_2,a_2).
\]

\noindent \textbf{Cognitive marker $c_2$:} refers to a political-legal concept, stressing institutional change and rational governance.  
\[
c_2 = \text{``institutional change toward democracy''}.
\]

\noindent \textbf{Affective marker ($a_2$):} relatively weak emotional tone; may evoke mild hope or optimism, but primarily cognitive.  
\[
a_2 = \text{``mild hope / optimism''}.
\]

\noindent \textbf{Representation in ${\cal H}_M$:}
\[
|m_2\rangle = |c_2\rangle \otimes |a_2\rangle 
= |\text{democratic reform}\rangle \otimes |\text{mild hope/optimism}\rangle.
\]
The affective marker component of this mental marker is very weak, so it can be treated as a cognitive marker.

\paragraph{Example 3: ``urgent danger''}  

\[
m_3 = (c_3,a_3).
\]

\noindent \textbf{Cognitive marker $c_3$:} signals a threat scenario, orienting attention toward risk and defensive action.  
\[
c_3 = \text{``perception of imminent threat''}.
\]
\noindent \textbf{Affective marker $a_3$:} strong emotional charge of fear, anxiety, and urgency.  
\[
a_3 = \text{``fear / anxiety / urgency''}.
\]

\noindent \textbf{Representation in ${\cal H}_M$:}
\[
|m_3\rangle = |c_3\rangle \otimes |a_3\rangle 
= |\text{imminent threat}\rangle \otimes |\text{fear/urgency}\rangle.
\]
The affective marker component of this mental marker is essentially stronger than its cognitive component, 
so it can be treated as an affective marker.

This way can construct  a dictionary of mental markers decomposed into their 
$(c,a)$ components. In SLT affective markers dominate as sources of social energy.

\section{Entanglement: Quantum and Quantum-like, Inter- and Intra-system} 

Entanglement represents one of the most profound nonclassical features of quantum theory, characterizing correlations that cannot be explained by local hidden variables. This is the standard statement which can be found in many textbooks and articles. But, typically one forgets to highlight that this statement is about non-contextual theories. By taking into account contextuality which is in turn fundamentally connected with incompatibility of observables (Bohr's principle of complementarity) we can decouple entanglement from nonlocality and in this way essentially demystify it (\cite{contextuality}, see also section \ref{ENTINC}).  

In modern applications---ranging from condensed matter to quantum-like cognition and social bahavioral modeling---it becomes crucial to distinguish \emph{entanglement within a system} (intra-system entanglement) from \emph{entanglement between distinct systems} (inter-system entanglement).

\subsection{Formal Definition of Entanglement}

\paragraph{Entanglement for Pure States}

Consider a quantum system with Hilbert state space $\mathcal{H}$ endowed with tensor ptoduct structure,   
$\mathcal{H} = \mathcal{H}_1 \otimes \mathcal{H}_2$. A pure state $|\psi\rangle \in \mathcal{H}$ is 
called \textbf{separable} if it can be written as a product state:
\[
|\psi\rangle = |\psi_1\rangle \otimes |\psi_2\rangle,
\]
where $|\psi_1\rangle \in \mathcal{H}_1$ and $|\psi_2\rangle \in \mathcal{H}_2$.  

If $|\psi\rangle$ cannot be expressed in this form, it is called \textbf{entangled}. For instance, the singlet state (one of the Bell states)   
\begin{equation}
\label{BBB}
|\Phi^- \rangle= (|-\rangle |+\rangle - |+\rangle |-\rangle)/ \sqrt{2}. 
\end{equation}
is a paradigmatic example of an entangled pure state.

\paragraph{Entanglement for Mixed States}

For a general mixed state described by a density operator $\rho$ on $\mathcal{H} = \mathcal{H}_1 \otimes \mathcal{H}_2$, the state is \textbf{separable} if it can be written as a convex combination of product states:
\[
\rho = \sum_i p_i \, \rho_1^{(i)} \otimes \rho_2^{(i)}, \quad p_i \ge 0, \quad \sum_i p_i = 1,
\]
where $\rho_1^{(i)}$ and $\rho_2^{(i)}$ are density operators on $\mathcal{H}_1$ and $\mathcal{H}_2$, respectively.  

If no such decomposition exists, $\rho$ is \textbf{entangled}. Mixed-state entanglement can be quantified using various measures such as the \emph{entanglement of formation}, \emph{negativity}, or \emph{concurrence}.

\paragraph{Inter-System Entanglement}

\textbf{Definition.} Inter-system entanglement concerns correlations between two or more distinct systems, each with its own Hilbert space. The composite state lives in $\mathcal{H}_1 \otimes \mathcal{H}_2$. A canonical example is the Bell state
$|\Phi^-\rangle,$ which entangles two distinct particles.

\textbf{Physical and informational meaning.} Inter-system entanglement enables nonlocal correlations and quantum communication phenomena such as teleportation and quantum key distribution. In social behavioral or cognitive analogs, it can represent shared meaning, mutual synchronization, or coordinated decisions across different agents or groups.

In physics inter-system entanglement is typically associated with ``quantum nonlocality'' - spooky action at the distance. 
As we have already pointed out, inter-system entanglement can be decoupled from such mystical nonlocality and it can be interpreted as 
a mathematical representation of contextuality-incompatibility (section \ref{ENTINC}). Thus by speaking about entanglement between states of social atoms, 
we don't refer to superluminal signaling between them. We refer to nonclassical correlations between them, nonclassicality means that it is impossible to describe all such correlations by a single Kolmogorov \cite{K} probability space.  

\paragraph{Intra-System Entanglement}

\textbf{Definition.} Intra-system entanglement refers to quantum correlations between degrees of freedom within a single system. For example, in an atom, the spin and orbital angular momentum of an electron can be entangled:
\[
|\psi\rangle = \frac{1}{\sqrt{2}}\big(|\uparrow\rangle_{\text{spin}}|1\rangle_{\text{orb}} + |\downarrow\rangle_{\text{spin}}|0\rangle_{\text{orb}}\big).
\]
This type of entanglement does not connect spatially separated entities but rather internal components of one entity \cite{IE1,IE2,IE3,IE4,IE5}.

\textbf{Conceptual role.} Intra-system entanglement underlies internal coherence, structure formation, and information integration. In quantum biology, such entanglement may contribute to efficient energy transfer in photosynthetic complexes. In quantum-like cognitive models, it can analogously describe correlations between internal cognitive states or conceptual dimensions within one agent.

\paragraph{Conceptual Comparison}

\begin{center}
\begin{tabular}{|l|l|l|}
\hline
\textbf{Aspect} & \textbf{Intra-System Entanglement} & \textbf{Inter-System Entanglement} \\
\hline
Subsystems & Internal degrees of freedom of one system & Distinct systems \\
\hline
Typical setting & Spin--orbit coupling & Bell pairs, multipartite states \\
\hline
Spatial locality & Local & Often nonlocal \\
\hline
Function & Internal coherence, self-organization & Correlation, communication, coordination \\
\hline
Social analogy  & Beliefs' coherence for one agent & States' synchronization across agents \\
\hline
\end{tabular}
\end{center}

\subsection{Contextuality-Incompatibility-Entanglement Triad}
\label{ENTINC}

To clarify the meaning of entanglement (both inter- and intra-system), consider the four-dimensional state space (two qubit space) and the singlet state $|\Phi^- \rangle$. It is entangled by definition, but what does this mean in physical terms? Consider two dichotomous observables $A$ and $B.$ If these obervables are compatible, then they can be jointly measurable, and the joint probability distribution  $P(A=\alpha, B=\beta|\Phi^-), \alpha, \beta= \pm 1,$ is well defined. In the mathematical formalism, observables are represented by commuting Hermitian operators that are defined by the same symbols; compatibility is expressed as commutativity of the operators $[A,B]=0.$   For each such pair, this is the classical probability distribution. Quantumness associated with the singlet state becomes visible if, instead one pair of compatible observables $(A,B),$ we consider a few pairs of compatible observables 
$(A_i,B_j),$ say $i,j=1,2,$ such that  $A$-observables $(A_1,A_2)$ as well $B$-observables $(B_1,B_2)$ are {\it incompatible}. In the operator formalism this means,      
\begin{equation}
\label{INC1}
[A_i, B_j] = 0,
\end{equation}
\begin{equation}
\label{INC2}
[A_1, A_2] \neq 0, \quad [B_1, B_2] \neq 0.
\end{equation}
Again for each pair of compatible observables we can define the joint probability distribution 
$P(A_i=\alpha, B_j=\beta|\Phi^-), \alpha, \beta= \pm 1.$ Each of them is classical. But it is impossible to describe all  
these probability distributions, $i,j=1,2,$ within the same Kolmogorov \cite{K} probability space. 

This can be proved with the aid of the well-known CHSH inequality \cite{Bell}. We note that constraints (\ref{INC1}) and (\ref{INC2}) formalize the CHSH-framework. In the quantum setting, without reference to hidden-variable theories, consider a pure quantum state $|\psi\rangle$ and the linear combination of correlations
\begin{equation}
\label{INC3}
\langle S \rangle_\psi = \langle A_1 B_1 \rangle_\psi + \langle A_1 B_2 \rangle_\psi + \langle A_2 B_1 \rangle_\psi - \langle A_2 B_2 \rangle_\psi.
\end{equation}
This can also be treated as the expectation value of the Bell operator:
\begin{equation}
\label{INC4}
S = A_1 B_1 + A_1 B_2 + A_2 B_1 - A_2 B_2.
\end{equation}
The CHSH inequality reads
\begin{equation}
\label{INC5}
|\langle S \rangle_\psi| \leq 2.
\end{equation}
As shown in \cite{GetRid}, the incompatibility constraint (\ref{INC2}) is a necessary condition for its violation for some state $|\psi\rangle$, and in the presence of a tensor product structure, it is also a sufficient condition for violation by some entangled state. Thus, Bell-type inequalities connect incompatibility and entanglement — {\it the incompatibility-entanglement link}.

{\it Contextuality} can be generally defined as the dependence of the outcomes of an observable on the experimental context — the full set of physical conditions under which the measurement is performed. This aligns with Bohr's view, connecting contextuality to the existence of incompatible observables and the principle of complementarity. In \cite{QL25} this framework was formalized as the principle of contextuality-complementarity. Bell adopted a more restricted view: a measurement context for observable $A$ is reduced to its joint measurement with an auxiliary observable $B$ compatible with $A$. Contextuality is expected only if observables $B_1$ and $B_2$ are incompatible. If $B_1$ and $B_2$ are compatible, a joint probability distribution exists and a classical, noncontextual description is possible. Otherwise, only pairwise distributions $p_{AB_i}, i=1,2$ exist, and contextuality tests reduce to violations of Bell-type inequalities.

This establishes the {\it Contextuality-Incompatibility-Entanglement Triad}. It is particularly useful for quantum-like modeling (QLM), especially in cognition and decision-making. In this context, contextuality implies that the outcome of a cognitive evaluation (a ``measurement'') depends on the context provided by other simultaneously evaluated markers. Contextuality, often expressed through the violation of Bell-type inequalities, is linked to the existence of incompatible mental markers and their entanglement. Because of this incompatibility, joint probability distributions for all markers cannot be defined classically. The entangled state $|\psi\rangle$ mathematically captures correlations between these contextual outcomes.  

For Social Laser Theory (SLT), this implies that a \textbf{collective social excitation} corresponds to a macroscopic state of inter-system contextuality, in which individual agents' responses are synchronized not by external force but by the shared contextual field of the information (infon) environment.

In QCDM framework the Bell inequality were employed both theoretically and experimentally in a few works \cite{Aerts1,Conte,BR1,Aerts2,QL3,Bruza,BR2,Behti,Gallus1,Gallus2,PK25,BR3,Aerts3}. 

\section{Illustrative Examples of Entangled Mental Markers}
\label{ENTEXP}

In contrast to the factorized forms considered previously, some mental markers cannot be represented as a simple tensor product
$|c\rangle \otimes |a\rangle$. Instead, their cognitive and affective components are \emph{entangled}, i.e., strongly correlated 
such that activation of one component determines or constrains the other. A standard mathematical criterion for entanglement is the Schmidt rank:
if the Schmidt decomposition (see Appendix C)
\[
|\psi_m\rangle = \sum_{i=1}^r \lambda_i\, |c_i\rangle \otimes |a_i\rangle, \qquad \lambda_i>0,\ \sum_i \lambda_i^2=1,
\]
has \emph{Schmidt rank} $r>1$, the state is entangled. For $r=1$, it reduces to a simple product state.

\paragraph{Example 4: Cognitive Dissonance (The Smoking Paradox)}
The tension of dissonance is modeled as an entangled state between health knowledge and immediate pleasure:
\[ |\psi_{\text{diss}}\rangle = \alpha |c_{\text{danger}}\rangle \otimes |a_{\text{anxiety}}\rangle + \beta |c_{\text{pleasure}}\rangle \otimes |a_{\text{relief}}\rangle \]
Here, the thought of danger is contextually locked to anxiety. The system cannot be in a state of "knowing danger" while "feeling relief" simultaneously.

\paragraph{Example 5: Brand Loyalty (Lovemarks)}
In consumer psychology, a product category becomes entangled with a deep identity marker:
\[ |\psi_{\text{brand}}\rangle = \frac{1}{\sqrt{2}}\big( |c_{\text{utility}}\rangle \otimes |a_{\text{neutral}}\rangle + |c_{\text{identity}}\rangle \otimes |a_{\text{devotion}}\rangle \big) \]
This explains why technical flaws ($c_{\text{utility}}$) are ignored by loyalists; the affective "branch" of the state dominates the subjective experience.

\paragraph{Example 6: Stereotype Threat}
This models the entanglement between a cognitive task and the fear of social evaluation:
\[ |\psi_{\text{threat}}\rangle = \sqrt{0.8}|c_{\text{task}}\rangle \otimes |a_{\text{anxiety}}\rangle + \sqrt{0.2}|c_{\text{logic}}\rangle \otimes |a_{\text{flow}}\rangle \]
The high weight of the anxiety branch demonstrates how social excitations (inter-system) induce performance-reducing entanglement within the individual (intra-system).

\paragraph{Example 7: ``protect our children from invasion''}  
\[
|\psi_{m_4}\rangle = \frac{1}{\sqrt{2}}\big(
|c_1\rangle \otimes |a_1\rangle 
+ 
|c_2\rangle \otimes |a_2\rangle
\big),
\]
where
\[
c_1 = \text{``perception of imminent invasion''}, \quad
a_1 = \text{``fear / anger''},
\]
\[
c_2 = \text{``protection of children and family''}, \quad
a_2 = \text{``urgent solidarity / care''}.
\]
Here, the emotional reaction (fear vs. protective solidarity) is inseparably tied to the cognitive frame (invasion vs. care). 
The Schmidt rank of this state is $r=2$, indicating entanglement.

\paragraph{Example 8: ``revolution for justice''}  
\[
|\psi_{m_5}\rangle = 
\sqrt{\tfrac{2}{3}}\,
|c_\mathrm{revolt}\rangle \otimes |a_\mathrm{anger}\rangle
+
\sqrt{\tfrac{1}{3}}\,
|c_\mathrm{justice}\rangle \otimes |a_\mathrm{hope}\rangle,
\]
where
\[
c_\mathrm{revolt} = \text{``uprising against injustice''}, \quad
a_\mathrm{anger} = \text{``anger / outrage''},
\]
\[
c_\mathrm{justice} = \text{``vision of fair order''}, \quad
a_\mathrm{hope} = \text{``hope / anticipation''}.
\]

This marker represents a mixture of destructive (angry revolt) and constructive (hopeful justice) channels. 
The unequal Schmidt coefficients $(\sqrt{2/3},\,\sqrt{1/3})$ reflect a stronger contribution from the anger-driven narrative.

\paragraph{Example 9: ``fight climate collapse''}  
\[
|\psi_{m_6}\rangle = 
\frac{1}{\sqrt{2}}\big(
|c_\mathrm{catastrophe}\rangle \otimes |a_\mathrm{fear}\rangle
+
|c_\mathrm{action}\rangle \otimes |a_\mathrm{determination}\rangle
\big),
\]
where
\[
c_\mathrm{catastrophe} = \text{``imminent climate catastrophe''}, \quad
a_\mathrm{fear} = \text{``fear / anxiety''},
\]
\[
c_\mathrm{action} = \text{``urgent mitigation measures''}, \quad
a_\mathrm{determination} = \text{``determination / resolve''}.
\]

Here, fear is directly coupled to catastrophic imagery, while determination is coupled to active mitigation, 
making the state entangled across fear and action narratives.

In Appendix D we consider toy numerical simulation for entanglement between cognitive and affective components of  
mental markers.

\subsection{Example of Incompatibility-Contextuality-Entanglement Triade for mental markers}

To illustrate the quantum-like structure of mental markers, consider an individual agent exposed to a socially relevant message or situation. We define two cognitive and two affective markers, each dichotomous (\(\pm 1\)), with both a psychological interpretation and a Hilbert-space representation.

\paragraph{Cognitive markers.} 
Let the cognitive Hilbert space be \(\mathcal{H}_C \cong \mathbb{C}^2\).
Consider two cognitive markers:
\begin{itemize}
    \item \(c_1 = \pm 1\) — \textit{Analytical evaluation}: 
    \[
    +1: \text{the situation is logically justified}, \quad 
    -1: \text{the situation is logically unjustified}.
    \]
    \item \(c_2 = \pm 1\) — \textit{Normative framing}: 
    \[
    +1: \text{the situation aligns with social norms}, \quad 
    -1: \text{the situation violates social norms}.
    \]
\end{itemize}

In quantum-like terms, these are represented by Pauli operators:
\[
c_1 = \sigma_z \otimes I_A, \qquad  c_2 = \sigma_x \otimes I_A.
\]
Since \([c_1, c_2] = 2 i \sigma_y \otimes I_A \neq 0\), \(c_1\) and \(c_2\) are incompatible. Psychologically, this models how focusing on logical analysis (\(c_1\)) can shift the context of normative judgment (\(c_2\)) and vice versa.

\paragraph{Affective markers.} 
Let the affective Hilbert space be \(\mathcal{H}_A \cong \mathbb{C}^2\). Define:
\begin{itemize}
    \item \(a_1 = \pm 1\) — \textit{Valence}: 
    \[
    +1: \text{positive emotional valence}, \quad 
    -1: \text{negative emotional valence}.
    \]
    \item \(a_2 = \pm 1\) — \textit{Arousal}: 
    \[
    +1: \text{high emotional arousal}, \quad 
    -1: \text{low emotional arousal}.
    \]
\end{itemize}
Quantum-like representation:
\[
a_1 = I_C \otimes \frac{\sigma_z + \sigma_x}{\sqrt{2}}, \qquad 
a_2 = I_C \otimes \frac{\sigma_z - \sigma_x}{\sqrt{2}}.
\]
The operators $a_1$ and $s_2$ are incompatible. This reflects how high arousal can disrupt valence evaluation, while calm states allow more differentiated emotional judgment.

We point out that each cognitive operator $c_i$ commutes with each affective operator $a_j$. 
Psychologically, this means an agent can simultaneously evaluate both \textit{what is thought} (logical and normative aspects) and \textit{how it feels} (valence and arousal).

\subsubsection{Interpretation: incompatibility vs. cross-compatibility} 
\begin{itemize}
    \item Incompatibility within the cognitive sector models competing cognitive frames (analytical vs. normative judgment).
    \item Incompatibility within the affective sector models competing emotional dimensions (valence vs. arousal).
    \item Compatibility between cognitive and affective markers allows joint cognition–emotion evaluation.
    \item Intra-system entanglement arises when specific cognitive frames become inseparably linked to particular affective states, e.g., a norm violation (\(c_2=-1\)) coupled with high arousal and negative valence (\(a_1=-1, a_2=+1\)).
\end{itemize}

Compatibility implies that the pairs $(c_i, a_j)$ admit well-defined joint eigenstates and generate legitimate product bases in the composite Hilbert space
\[
\mathcal{H}_M = \mathcal{H}_C \otimes \mathcal{H}_A.
\]
The four bases will be  considered in the following analysis:
 the $(c_1, a_1)$ basis, consisting of joint eigenstates of $c_1=\sigma_z$ and $a_1=(\sigma_z+\sigma_x)/\sqrt{2}$;
 the $(c_1, a_2)$ basis, consisting of joint eigenstates of $c_1=\sigma_z$ and $a_2=(\sigma_z-\sigma_x)/\sqrt{2}$;
 the $(c_2, a_1)$ basis, consisting of joint eigenstates of $c_2=\sigma_x$ and $a_1=(\sigma_z+\sigma_x)/\sqrt{2}$;
 the $(c_2, a_2)$ basis, consisting of joint eigenstates of $c_2=\sigma_x$ and $a_2=(\sigma_z-\sigma_x)/\sqrt{2}$.
Each basis (or a pair of commuting operators acting in the components of the tensor product) determines a decision context. 
The same quantum state $|\psi\rangle \in \mathcal{H}_M$ can be expanded w.r.t. to each of these bases; each such expansion 
is interpreted as context selection.

This example demonstrates how real cognitive and affective markers can be represented mathematically using Pauli matrices, making the Hilbert-space formalism concrete while retaining psychological interpretability; see Table \ref{tab:markers}.

\begin{table}[h!]
\centering
\caption{Cognitive and affective markers: psychological interpretation and quantum-like representation}
\begin{tabular}{|c|c|c|c|}
\hline
\textbf{Marker} & \textbf{Values} & \textbf{Psychological meaning} & \textbf{Pauli matrices} \\
\hline
$c_1$ & $\pm 1$ & Analytical evaluation: logically justified/unjustified & $\hat c_1 = \sigma_z \otimes I_A$ \\
$c_2$ & $\pm 1$ & Normative framing: aligns with/violates norms & $\hat c_2 = \sigma_x \otimes I_A$ \\
\hline
$a_1$ & $\pm 1$ & Valence: $+1$ positive, $-1$ negative & $\hat a_1 = I_C \otimes \frac{\sigma_z + \sigma_x}{\sqrt{2}}$ \\
$a_2$ & $\pm 1$ & Arousal: $+1$ high, $-1$ low & $\hat a_2 =  I_C \otimes \frac{\sigma_z + \sigma_x}{\sqrt{2}}$ \\
\hline
\end{tabular}
\label{tab:markers}
\end{table}

\subsubsection{Contextual interpretation of a quantum state}
\label{BASIS} 

This illustrative example shows that a quantum state $|\psi\rangle$ as the normalized vector of Hilbert vstate space is just a mathematical symbol. Its cognitive (as well as physical) interpretation is possible only after selection of the corresponding context - an orthonormal basis (or a pair of operators). Saying that two agents have the same mental state $|\psi\rangle$ has no meaning without pointing to contexts used by these agents.   

\subsection{Contextual psychological representations of singlet-state}

We consider a singlet-like intra-system entangled mental marker state:
\begin{equation}
|\Psi^-\rangle = \frac{1}{\sqrt{2}} \Big( |c_1=+, a_1=-\rangle - |c_1=-, a_1=+\rangle \Big),
\end{equation}
where $c_1$ denotes a cognitive marker (analytical evaluation) and $a_1$ a rotated affective marker combining valence and arousal. The state exhibits perfect anti-correlation between cognitive judgment and affective response along the $c_1, a_1$ axes.

\paragraph{1. Expansion in the $(c_1, a_1)$ basis} 
The state is already expressed in the $a_1$ eigenbasis:
\[
|\Psi^-\rangle = \frac{1}{\sqrt{2}} \big( |c_1=+, a_1=-\rangle - |c_1=-, a_1=+\rangle \big).
\]

\textbf{Psychological interpretation:} 
Cognitive evaluation and affective response are perfectly anti-correlated: a logically justified judgment ($c_1=+$) triggers a negative valence response ($a_1=-$), whereas a logically unjustified judgment ($c_1=-$) triggers positive valence ($a_1=+$).  
This captures the essence of cognitive-emotional tension, analogous to cognitive dissonance (cf. \cite{PolinaCD}).  
At the collective level this is unstable society.

Now consider the state
\[
|\Psi^-\rangle = \frac{1}{\sqrt{2}} \big( |c_1=+, a_1=+\rangle - |c_1=-, a_1=-\rangle \big).
\]
\textbf{Psychological interpretation:} 
- Cognitive evaluation and affective response are perfectly correlated: a logically justified judgment ($c_1=+$) triggers a positive valence response ($a_1=+$), whereas a logically unjustified judgment ($c_1=-$) triggers negative valence ($a_1=-$). In this state agents enjoy analytic evaluation of problems. At the collective level this is a ``rational society''.    

\paragraph{2. Expansion in the $(c_1, a_2)$ basis}
The $a_1$ eigenstates in the $a_2$ basis are:
\[
|a_1=+\rangle = \frac{1}{\sqrt{2}}(|a_2=+\rangle + |a_2=-\rangle), \quad
|a_1=-\rangle = \frac{1}{\sqrt{2}}(-|a_2=+\rangle + |a_2=-\rangle).
\]

Then the singlet expands as:
\[
|\Psi^-\rangle = \frac{1}{2} \Big[ -(|c_1=+\rangle + |c_1=-\rangle)|a_2=+\rangle + (|c_1=+\rangle - |c_1=-\rangle)|a_2=-\rangle \Big].
\]

\textbf{Psychological interpretation:} Expanding $|\Psi^-\rangle$ in the $(c_1,a_2)$ basis reflects a change of context, combining analytical cognition ($c_1$) with arousal-based affect ($a_2$). The state does not assign definite values to either marker but defines the questions used to interpret the mental state. Analytical judgment appears as a coherent superposition of alternatives, showing that, in arousal-relevant contexts, cognition is indeterminate and context-dependent. At the collective level, shared adoption of this context aligns cognitive states, enabling coherent responses even without uniform beliefs.

\paragraph{3. Expansion in the $(c_2, a_1)$ basis}
Rotate the cognitive axis: 
\[
|c_2=+\rangle = \frac{|c_1=+\rangle + |c_1=-\rangle}{\sqrt{2}}, \quad 
|c_2=-\rangle = \frac{|c_1=+\rangle - |c_1=-\rangle}{\sqrt{2}}.
\]

Then:
\[
|\Psi^-\rangle = \frac{1}{2} \Big[ |c_2=+\rangle (-|a_1=+\rangle + |a_1=-\rangle) + |c_2=-\rangle (|a_1=+\rangle + |a_1=-\rangle) \Big].
\]

\textbf{Psychological interpretation:}
Normative framing or reframing of the situation ($c_2$) introduces partial superpositions in affective response.  
Anti-correlation between cognition and emotion is now ``tilted'', meaning that normative considerations alter emotional reactions without eliminating the underlying cognitive-emotional tension.  
This illustrates frame-dependence in judgment: the same affective input can lead to different cognitive interpretations depending on cognitive framing, a phenomenon widely observed in behavioral economics and social psychology.  
At the collective level, this predicts that changing the framing of social information can reshuffle population-level responses, potentially amplifying or dampening social coherence.

\paragraph{4. Expansion in the $(c_2, a_2)$ basis}
Finally, expressing both rotations simultaneously:

\[
|\Psi^-\rangle = \frac{1}{2} \Big[ -(|c_2=+\rangle + |c_2=-\rangle)|a_2=+\rangle + (|c_2=+\rangle - |c_2=-\rangle)|a_2=-\rangle \Big].
\]

\textbf{Psychological interpretation:}ctive behavior.
Within this context, the quantum state shows that arousal-sensitive evaluation structures normative cognition as a superposition of potential alternatives rather than fixed judgments. Individually, normative evaluation is non-classical and context-dependent; collectively, shared adoption of the $(c_2,a_2)$ context aligns cognitive–affective states, allowing populations to respond coherently to stimuli. In Social Laser Theory terms, this maximal potentiality primes the population for information pumping, enabling rapid, coordinated collective behavior without uniform beliefs or explicit coordination.

\paragraph{Summary}
Representation of a quantum state $|\Psi^-\rangle$ bases $(c_1, a_1), (c_1, a_2), (c_2, a_1), (c_2, a_2)$ correspond to selection of four different contexts in which an agent can take decisions, this  illustrates:- Cognitive judgments and affective markers are ``entangled in a context-dependent way''.  - Different basis choices correspond to ``different cognitive-emotional perspectives, a hallmark of QLM of decision-making.- This beautifully represents how context reshapes the ``space of cognitive possibilities''.
  - At the collective level, these intra-system correlations can propagate as inter-agent entanglement, driving social coherence and collective behavioral excitations.

\subsubsection{CHSH correlation}

The CHSH operator is
\[
S = c_1 a_1 + c_1 a_2 + c_2 a_1 - c_2 a_2.
\]
Using the singlet-like state properties:
we compute
$
\langle c_1 a_1 \rangle=\langle c_1 a_2 \rangle=\langle c_2 a_1 \rangle = -\frac{1}{\sqrt{2}}, \;
\langle c_2 a_2 \rangle = +\frac{1}{\sqrt{2}}.
$
Thus,
\[
\langle S \rangle = - 2\sqrt{2}.
\]

\textbf{Interpretation:} 
This maximal violation of the classical CHSH bound ($|\langle S \rangle| = 2\sqrt{2}$) demonstrates that the joint cognitive-affective state cannot be represented by a single global probability distribution. Instead, the correlations are context-dependent: the outcome of a cognitive evaluation (e.g., $c_1$) is inextricably linked to the affective context ($a_1$ or $a_2$) in which it is measured.At the individual level, the violation proves that the agent does not possess a pre-scripted list of responses for all possible questions. Rather, the cognitive judgment and affective response emerge only within a specific, chosen context. At the collective level, this implies that the Social Laser does not rely on individual agents having uniform, fixed beliefs. Instead, coherence is achieved by synchronizing the contexts (the specific bases $(c_i, a_j)$) across the population. When a population shares an interpretational context, their quantum-like mental states become capable of the high-strength correlations necessary for mass mobilization or rapid shifts in public opinion, even if their individual starting points are indeterminate.¨

\subsection{Impact of Affective Markers on Social Lasing}
\label{Affect}

Affective markers play a central role in political mobilization, mass protests, and advertising aimed at shaping consumer behavior or lifestyles. These markers typically carry minimal semantic load but evoke strong emotions—anger, pride, fear, or even amusement—thereby enabling cost-effective and wide-reaching dissemination. Their primary communicative function lies in mobilizing affect rather than conveying propositional content.

From a semiotic perspective, affective markers operate predominantly on the level of connotation \cite{Barthes} and exemplify affective investment in signifiers  \cite{Laclau}, through which otherwise empty or floating signifiers gain mobilizing power. In communication theory, this aligns with low-information, high-impact messaging \cite{McCombs,Entman}, where emotional resonance, rather than informational depth, drives framing, agenda-setting, and discourse saturation. Habermas  \cite{Habermas}  warned that such strategic communication risks privileging manipulation over deliberation.

Empirical studies confirm these effects: viral advertising research shows that emotional tone strongly shapes attitudes and forwarding intentions \cite{Eckler}, while motivational and emotional triggers outperform rational appeals 
in predicting sharing behavior \cite{Dafonte-Gomez}. Recent reports for the UN Economic and Social Council \cite{Rachmad1,Rachmad2} highlight the ``viral'' and ``gimmick'' effects as catalysts for behavioral contagion and discourse alignment.

Within SLT, affective markers can be viewed as discrete ``quanta'' of social energy capable of exciting social atoms (individuals or groups) into higher levels of affective arousal. Repetition and amplification of such markers induce coherence across populations, aligning emotional phases and enabling collective mobilization. Thus, classical framing and agenda-setting mechanisms can be reinterpreted as structuring the energy landscape of discourse.

\section{Concluding Remarks}

In this paper, we have proposed QLM of mental markers within the broader framework of QCDM and the $I$-field approach (coupled to SLT). By focusing on the internal structure of content-carrying markers—rather than on large-scale collective dynamics—we have highlighted the fundamental role of contextuality, incompatibility, and entanglement in shaping human cognition under conditions of informational saturation. In particular, we have argued that cognitive and affective components of mental markers are not independent dimensions, but may form an intrinsically entangled structure. This intra-system entanglement provides a natural explanation for the contextual linkage between rational evaluation and emotional coloring observed in decision-making.

Several perspectives follow from this analysis. First, the proposed formalism opens the possibility of constructing experimentally testable models of cognitive–affective entanglement, for example through the analysis of order effects, framing manipulations, and interference-like patterns in judgment tasks. Second, the distinction between inter-system and intra-system entanglement invites further investigation of how individual-level marker structures scale to collective informational dynamics. Third, the quantum-field-inspired representation of the information environment suggests new ways of modeling how structured informational inputs shape internal cognitive states over time.

More broadly, this work contributes to the ongoing development of QLM as a unified theoretical language for describing nonclassical aspects of cognition. By clarifying the structural organization of mental markers, we hope to stimulate interdisciplinary dialogue between cognitive science, social psychology, semiotics, decision theory, social  mobilization, and mathematical modeling, and to encourage empirical studies capable of evaluating the explanatory power of the proposed framework. In Appendix B we discuss
 potential applications of our QLM to neurobiology and medicine. This line of thought can lead to further applications of QLM 
for teh diagnostics neurological deceases. 

\section*{Appendix A: Mental markers vs. semantic primitives, semiotics of passions, narrative sociology}

The cognitive marker $c$ may be aligned with Wierzbicka's theory of \emph{semantic primitives and
universals} \cite{Vezhbitskaya2022}, which provides a compact inventory of basic
conceptual atoms that serve as a basis for meaning representation. 

The affective marker $a$, in turn, resonates with Barthes' semiotics of connotation \cite{Barthes} and with Greimas'
``semiotics of passions'' \cite{GreimasFontanier2007}, where passions are modeled as
continuous meaning-entities linking embodied affect to semiotic form. 

Finally, the distribution and resonance of
mental markers across a population is embedded in macro-level discursive structures. Here,
the sociology of justification developed by Boltanski-Thevenot
\cite{BoltanskiThevenot2013} and the narrative sociology tradition
\cite{YarskayaSmirnova1997,Shcherbin2024} provide the framework that determines which
cognitive markers are foregrounded, how they are paired with affective markers, and how
mental markers achieve collective resonance and coherence.

In this way, semiotics and narrative sociology supply invention of mental markers in QCDM 
with a theoretical foundation: Wierzbicka's semantic atoms
define the space of cognitive markers, Barthes \cite{Barthes} and  Greimas-Fontanier  \cite{GreimasFontanier2007} clarify the role of affective
markers and their entanglements, and Boltanski--Thevenot \cite{BoltanskiThevenot2013} explain the higher-order frames
that regulate the circulation and amplification of mental markers.

\newpage
\section*{Appendix B: Neurobiological Analogues: Information Overload in Neural Networks} 

The preceding sections developed a quantum-informational model of mental markers under conditions of information overload, where cognitive agents respond to compact content labels rather than to the full semantic content of messages. A fundamental assumption of this framework is that information overload forces a compression of rich informational input into discrete $(E, m)$ quanta. We now show that this assumption has deep neurobiological analogues: several neurological conditions can be understood as disorders in which neural networks fail to compress, filter, or manage incoming information, leading to pathological system responses. These clinical parallels are not merely metaphorical, they share structural features with the $I$-field framework, including episodicity, threshold-dependent phase transitions, and the role of excitatory–inhibitory balance in maintaining contextual processing capacity. 

We examine five neurological conditions: migraine, epilepsy, autism spectrum disorder (ASD), attention deficit hyperactivity disorder (ADHD), and Alzheimer’s disease (AD). Each of which manifests a distinct mode of failure when neural information processing is overwhelmed. Together, they form a spectrum of network overload responses that illuminates the generality of the information-theoretic principles underlying our quantum-like model of mental markers. 

The most reproducible neurophysiological abnormality in migraine between attacks is a deficit of habituation to repeated sensory stimuli across virtually all modalities: visual, auditory, somatosensory, and nociceptive \cite{100, 101, 102}. Habituation is the brain’s fundamental mechanism for filtering redundant sensory input; Coppola and colleagues \cite{101} have characterized it as the mechanism that “protects the cortex against the overflow of inward information.” In migraineurs, this protective filter fails: instead of diminishing their response to repetitive stimulation, cortical evoked potentials show progressive potentiation \cite{103}. The result is a continuous accumulation of unfiltered sensory information. This habituation deficit has been linked to a dysrhythmic thalamocortical activity pattern driven by inadequate monoaminergic control \cite{101, 104}. The thalamus acts as a sensory gating structure, reducing the throughput of redundant signals to the cortex. When this gating fails, the cortex is exposed to information flows that exceed its metabolic processing capacity. Gross and colleagues \cite{105} have proposed an energy information mismatch theory of migraine: the attack is triggered when the energy demands of processing accumulated sensory information exceed the available metabolic supply. In this framework, the migraine attack itself (with its cortical spreading depression, trigeminal activation, and sensory shutdown) functions as a protective system reset, analogous to a circuit breaker that disconnects an overloaded electrical network. 

The connection to the $I$-field framework is direct. In our model, mental markers emerge precisely because agents under information overload cannot process full message content and must instead respond to compact labels. The migrainous brain fails at precisely this kind of neural compression: it cannot reduce rich sensory streams to manageable representations. The cycling nature of migraine with the habituation deficit building progressively before the attack and normalizing during it \cite{102} mirrors a dynamic of information accumulation and discharge that Social Laser Theory describes at the social level. The migraine attack can be interpreted as the neural system’s attempt to restore the capacity for marker formation by resetting its overloaded processing architecture. Schoenen \cite{106} explicitly linked the habituation deficit to a failure in ``information processing,'' proposing that deficient habituation reflects an inability of the sensory cortex to properly reduce its response to incoming data. Goadsby and colleagues \cite{107} have reframed migraine as a ``sensory processing disorder,'' emphasizing that the pathophysiology centers on how the brain handles sensory information rather than on vascular changes. This reconceptualization aligns the migraine brain with a system whose information compression mechanisms have broken down. 

If migraine represents a gradual protective shutdown in response to information overload, epilepsy represents its catastrophic counterpart: the collapse of differentiated neural processing into pathological hypersynchronization. Reflex epilepsies, comprising 4 to 7\% of all epilepsies \cite{108}, provide the clearest evidence that specific sensory inputs can directly overwhelm epilepsy-prone networks. In photosensitive epilepsy, flickering light at specific frequencies triggers seizures by driving cortical circuits beyond their capacity to maintain excitatory inhibitory $(E/I)$ balance \cite{109}. Ferlazzo and colleagues \cite{109} demonstrated that in reflex seizures, the triggering stimulus activates cortical networks in a way that exceeds their processing threshold, causing a transition from normal differentiated activity to pathological hypersynchrony. Crucially, the distinction between reflex and spontaneous epilepsy may be one of degree rather than kind. Irmen and colleagues \cite{110} have proposed a continuum hypothesis, arguing that many apparently spontaneous seizures are in fact triggered by intrinsic information-processing events: specific cognitive tasks, emotional states, or internal oscillatory patterns, that push vulnerable networks past their seizure threshold. Wolf and Bhatt \cite{111} have argued that the reflex mechanism in which a specific input overloads a susceptible network, may represent a general component of ictogenesis rather than a special category. This perspective aligns directly with the $I$-field model: both extrinsic stimuli (the information field) and intrinsic processing (within-agent cognitive dynamics) can generate overload. Modern network theories of epilepsy have moved beyond simple focus models to conceptualize seizures as emergent properties of brain network architecture \cite{112, 113}. Terry and colleagues \cite{114} emphasize that epilepsy is fundamentally a network-level disorder in which the architecture of neural connectivity determines vulnerability to hypersynchronization. In this view, a seizure represents the catastrophic loss of the tensor product structure that allows for differentiated, context-dependent processing across cognitive and affective degrees of freedom, precisely the structure formalized in our model as $H_M = H_C \otimes H_A.$ During a seizure, all neurons synchronize into a single mode, destroying the contextuality that allows for nuanced cognitive–affective processing. In the language of our framework, the seizure collapses the capacity for maintaining incompatible mental observables and their entangled correlations. 

Autism spectrum disorder (ASD) provides a third mode of information overload pathology, distinct from both migraine and epilepsy. Approximately 90\% of individuals with ASD experience sensory processing differences \cite{115}, and hyper- or hypo-reactivity to sensory input is now included as a diagnostic criterion in the DSM-5. Several theoretical frameworks converge on information overload as central to the autistic experience. The intense world theory \cite{116} proposes that increased local neural connectivity and plasticity lead to hyperperception, an excessive, unfiltered processing of sensory detail that makes the world overwhelming. The weak central coherence theory \cite{117} describes a locally focused processing style in which individuals process information in fine detail but have difficulty integrating it into global contextual representations. The Bayesian predictive coding account \cite{118} suggests that autistic perception is biased toward sensory error signals, with reduced attenuation of prediction errors, leading to a perceptual world that is noisy, unpredictable, and difficult to compress. Neurophysiologically, ASD is characterized by increased local cortical connectivity coupled with impaired long-range connectivity \cite{119, 120}. This architecture produces an information processing system that excels at capturing local detail but fails at the kind of hierarchical compression needed to reduce rich sensory streams to manageable representations, the neural equivalent of mental marker formation. Marco and colleagues \cite{121} showed that children with autism may rely more heavily on already overloaded attention and working-memory networks, such that when stimuli exceed capacity, the processing system fails. The deficit in sensory gating and habituation observed in ASD \cite{120} parallels the migraine habituation deficit, while the $E/I$ imbalance identified in ASD networks \cite{122} connects to the epilepsy literature. Indeed, the co-occurrence of epilepsy and ASD (approximately 20–30\% of individuals with ASD develop epilepsy) suggests shared vulnerability to network overload. For the I-field model, ASD illustrates what happens when the neural substrate for marker compression is atypical: the system processes too much raw information and too little compressed, label-like output. The monotropism account \cite{123} describes autistic attention as supporting few synchronous interests, each highly aroused and like a pattern that can be interpreted as the cognitive system’s adaptation to limited capacity for simultaneous marker processing. 

Attention deficit hyperactivity disorder (ADHD) offers yet another variant of information overload pathology. While ASD involves excessive bottom-up sensory processing, ADHD is primarily characterized by impaired top-down attentional control over incoming information \cite{124}. The ADHD brain has a reduced ability to filter out irrelevant stimuli, meaning it processes everything at once rather than selectively attending to relevant signals \cite{125}. This produces sensory overload not because the system is hyper-responsive (as in migraine) or hyper-connected locally (as in ASD), but because it fails to prioritize. Research by Friedman-Hill and colleagues \cite{124} suggests that distractibility in ADHD results from deficits in top-down cognitive control rather than from bottom-up sensory competition. ADHD children were most distractible with low-salience distractors and during easy tasks inconsistent with bottom-up filtering failure but consistent with deficient prefrontal capacity to suppress irrelevant input. The combination of impaired filtering with lower working memory capacity and slower processing speed \cite{125} means that the ADHD information processing system is both receiving more inputs and processing them less efficiently. Sensory processing problems are significantly more common in individuals with ADHD than in typically developing populations \cite{126}, with up to 60\% exhibiting symptoms of sensory processing disorder. In the I-field framework, ADHD can be understood as a disorder of marker selection: the system receives the full information field but cannot efficiently extract and respond to the relevant mental markers. Instead of compressing information into actionable (c, a) pairs, the ADHD system oscillates between all available markers without stable selection. This connects to the clinical observation that ADHD individuals often show both sensory-seeking behavior (insufficient marker activation, leading to stimulation-seeking) and sensory overload (excessive marker activation when the environment provides too many simultaneous excitations). 

Alzheimer's disease (AD) provides a fifth perspective on neural information overload, distinguished from the preceding conditions by its progressive and neurodegenerative character. Recent evidence indicates that network hyperexcitability including hypersynchrony, altered oscillatory rhythmic activity, and interneuron dysfunction and others, as a key mediator of cognitive decline in AD, not merely a consequence of neuronal loss \cite{127, 128}. In early-stage AD, soluble amyloid-$\beta (A\beta)$ disrupts excitatory–inhibitory balance, initially producing hippocampal and cortical hyperactivity before progressing to hypoactivity in later stages \cite{129, 130}. This hyperexcitability manifests as epileptiform activity, with AD patients having a significantly increased risk of seizures directly connecting AD to the epilepsy framework described above. Palop and Bhatt \cite{131} demonstrated that network hypersynchrony in AD arises in part from impaired inhibitory interneuron function, particularly of parvalbumin-positive (PV) interneurons that normally maintain gamma oscillatory activity essential for learning and memory. A multiscale computational model by van Nifterick and colleagues \cite{132} showed that microscale neuronal hyperexcitability in AD can explain the characteristic large-scale oscillatory slowing observed in MEG recordings of early-stage patients. De Haan and colleagues \cite{133} demonstrated that functional connectivity measures track E/I balance and could serve as biomarkers for network hyperexcitability. In the I-field framework, AD represents the progressive dissolution of the neural infrastructure needed for marker formation and contextual processing. As inhibitory networks degrade, the brain’s ability to maintain differentiated, context-dependent representations, the tensor product structure $H_C \otimes H_A$ with its contextual bases collapses progressively. The hyperexcitability phase can be interpreted as the system’s transitional state when $E/I$ balance is shifting but structure is not yet fully dissolved, analogous to the critical phase before a seizure in the epilepsy model. The eventual cognitive decline represents the final loss of marker-forming capacity. 

All five conditions share key features with the $I$-field model: 
\begin{itemize}
\item (i) the centrality of information processing capacity as a limiting factor; 
\item (ii) the role of excitatory–inhibitory balance in maintaining or losing contextual differentiation; 
\item (iii) threshold-dependent transitions between normal and pathological states; 
\item  (iv) the network-level (rather than focal) nature of the dysfunction. 
\end{itemize}
The neural habituation deficit in migraine corresponds to the failure of social agents to filter redundant information. The hypersynchronization of epilepsy corresponds to the coherent, undifferentiated collective response modeled in SLT. The autistic information processing style illustrates the cost of failing to compress sensory input into manageable markers. The ADHD filtering deficit parallels the selective attention failures that make agents vulnerable to information field excitations. And the progressive collapse in AD models the long-term degradation of the cognitive architecture needed for marker-based processing. These neurobiological parallels validate the core assumption of the $I$-field model: that information overload is not merely a subjective experience but a fundamental computational constraint on any system (biological or social) that must process structured input under finite resources. The quantum-like formalism developed in this paper provides a mathematical framework that can, in principle, accommodate all five modes of overload response, since the tensor product structure, contextuality, incompatibility, and entanglement it employs are precisely the features that are disrupted or collapsed in each pathological condition. 

\section*{Appendix C: Schmidt Decomposition}

{\bf Theorem.} 
Let ${\cal H}_C$ and ${\cal H}_A$ be finite-dimensional Hilbert spaces of dimensions 
$d_C$ and $d_A$, and let ${\cal H} = {\cal H}_C \otimes {\cal H}_A$ be their tensor product. 
For any pure state $|\psi\rangle \in {\cal H}$, there exist orthonormal bases 
$\{|c_i\rangle\}_{i=1}^{d_C}$ of ${\cal H}_C$ and 
$\{|a_i\rangle\}_{i=1}^{d_A}$ of ${\cal H}_A$, and a set of non-negative real coefficients
$\{\lambda_i\}_{i=1}^r$, with
\[
\lambda_i \ge 0,\qquad \sum_{i=1}^r \lambda_i^2 = 1,\qquad 
r \le \min(d_C,d_A),
\]
such that $|\psi\rangle$ can be written in the \emph{Schmidt form}
\begin{equation}
\label{eq:Schmidt}
|\psi\rangle = \sum_{i=1}^r \lambda_i\,|c_i\rangle\otimes|a_i\rangle.
\end{equation}

The integer $r$, called the \emph{Schmidt rank}, is uniquely determined by the state $|\psi\rangle$
and equals the rank of either reduced density matrix
$\rho_C = \mathrm{Tr}_A(|\psi\rangle\langle\psi|)$ or 
$\rho_A = \mathrm{Tr}_C(|\psi\rangle\langle\psi|)$.

Moreover, the nonzero coefficients $\lambda_i$ are the square roots of the nonzero eigenvalues of
$\rho_C$ and $\rho_A$, and the vectors $\{|c_i\rangle\}$ and $\{|a_i\rangle\}$ are the corresponding
orthonormal eigenvectors.

\section*{Appendix D: Toy Numerical Simulation: Mental Marker with Entangled Cognitive and Affective Components}

Consider a two-dimensional cognitive space spanned by orthogonal basis $\{|c_1\rangle,|c_2\rangle\}$ and a two-dimensional affective space spanned by orthogonal basis $\{|a_1\rangle,|a_2\rangle\}$.  
Take an entangled mental-marker state
\begin{equation}
\label{psi_ent}
|\psi\rangle = \lambda_1\,|c_1\rangle\otimes|a_1\rangle + \lambda_2\,|c_2\rangle\otimes|a_2\rangle,
\end{equation}
with coefficients 
$
\lambda_1=\sqrt{0.7}\approx 0.83666,\qquad \lambda_2=\sqrt{0.3}\approx 0.54772,
$
so that $\lambda_1^2+\lambda_2^2=1$.

\paragraph{Joint state (density matrix).}
The pure-state density operator is
$$
\rho=|\psi\rangle\langle\psi|
= \begin{pmatrix}
0.7 & 0 & 0 & \lambda_1\lambda_2\\[4pt]
0 & 0 & 0 & 0\\[4pt]
0 & 0 & 0 & 0\\[4pt]
\lambda_1\lambda_2 & 0 & 0 & 0.3
\end{pmatrix},
$$
written in the ordered product basis $\{|c_1a_1\rangle,|c_1a_2\rangle,|c_2a_1\rangle,|c_2a_2\rangle\}$.  
(Only the entries corresponding to $|c_1a_1\rangle$ and $|c_2a_2\rangle$ are occupied because of the chosen Schmidt form.)

\paragraph{Marginal (subjective) probabilities --- Born rule.}
The probability that a social atom assigns cognitive marker $c_i$ is
$$
p(c_i)=\mathrm{Tr}\big[(|c_i\rangle\langle c_i|\otimes I_a)\,\rho\big].
$$
These reduce to
$$
p(c_1)=\lambda_1^2=0.7,\qquad p(c_2)=\lambda_2^2=0.3.
$$
Similarly for affective markers:
$$
p(a_1)=\lambda_1^2=0.7,\qquad p(a_2)=\lambda_2^2=0.3.
$$
\paragraph{Conditional probabilities.}
The nonzero joint probabilities are
$$
p(c_1,a_1)=0.7,\qquad
p(c_2,a_2)=0.3,
$$
and all cross terms vanish:
$$
p(c_1,a_2)=p(c_2,a_1)=0.
$$
The conditional probabilities are obtained via
$$
p(c_i|a_j)=\frac{p(c_i,a_j)}{p(a_j)},\qquad
p(a_j|c_i)=\frac{p(c_i,a_j)}{p(c_i)}.
$$
Since
$$
p(c_1)=p(a_1)=0.7,\qquad p(c_2)=p(a_2)=0.3,
$$
we find
$$
p(c_1|a_1)=\frac{0.7}{0.7}=1,\qquad
p(c_2|a_2)=\frac{0.3}{0.3}=1,
$$
and for the cross pairs,
$$
p(c_1|a_2)=p(c_2|a_1)=0,\qquad
$$
In the same way we get that 
$$
p(a_i|c_i)=1, i=1,2, \mbox{and} p(a_i|c_j)=0, i \not=j.
$$
Thus, the cognitive and affective components are perfectly correlated: observing $c_i$ implies $a_i$ and vice verse. 

\paragraph{Reduced density matrices.}
Tracing out the affective subsystem gives the cognitive reduced state
$$
\rho_c=\mathrm{Tr}_a(\rho)=\begin{pmatrix}0.7 & 0\\[4pt]0 & 0.3\end{pmatrix}
\quad\text{(in the }\{|c_1\rangle,|c_2\rangle\}\text{ basis).}
$$
Similarly, the affective reduced state is
$$
\rho_a=\mathrm{Tr}_c(\rho)=\begin{pmatrix}0.7 & 0\\[4pt]0 & 0.3\end{pmatrix}
\quad\text{(in }\{|a_1\rangle,|a_2\rangle\}\text{).}
$$

\paragraph{Entanglement entropy (von Neumann entropy).}
For a pure bipartite state the entanglement is given by the von Neumann entropy of either reduced state:
\[
S(\rho_c) = -\sum_{i=1}^2 \lambda_i^2 \log_2(\lambda_i^2).
\]
With $\lambda_1^2=0.7$ and $\lambda_2^2=0.3$,
$$
S(\rho_c) = -\big(0.7\log_2 0.7 + 0.3\log_2 0.3\big)
\approx 0.8813 .
$$
Interpretation: $S\approx 0.88$ bits indicates a substantial (but not maximal) entanglement / cognitive–affective coupling.

\paragraph{Comparison: factorized (product) state.}
For contrast, consider the product state
\[
|\phi\rangle = (\sqrt{0.7}\,|c_1\rangle + \sqrt{0.3}\,|c_2\rangle)\otimes
(\sqrt{0.5}\,|a_1\rangle + \sqrt{0.5}\,|a_2\rangle).
\]
Here the joint probabilities factorize:
$$
p(c_i,a_j)=p(c_i)\,p(a_j),
$$
and the reduced-state entropies are independent; the entanglement entropy $S(\rho_c)=0$ (no entanglement). In particular, detecting $a_1$ gives no information about which cognitive marker $c_i$ is realized, unlike the entangled case above.

\paragraph{Practical remark for SLT / QLM.}
The entangled toy state (\ref{psi_ent}) models a mental marker where cognitive frames and emotions are tightly coupled: the activation of a specific affective tone (say, fear) carries information about the cognitive framing (perception of invasion) and vice versa. Such markers can produce stronger coherence and higher lasing efficiency in SLT because single excitations project onto 
correlated    cognitive–affective responses.

\paragraph{Non-perfect correlation: toy example.}

Consider a slightly non-perfectly-correlated two-by-two state with small cross terms,
\[
|\psi\rangle = \sqrt{0.6}\,|c_1\rangle|a_1\rangle
+ \sqrt{0.1}\,|c_1\rangle|a_2\rangle
+ \sqrt{0.1}\,|c_2\rangle|a_1\rangle
+ \sqrt{0.2}\,|c_2\rangle|a_2\rangle,
\]
so that the squared moduli of the coefficients sum to unity.

\paragraph{Joint probabilities (Born rule).}
The joint probabilities are the squared moduli of the amplitudes in the product basis:
\[
\begin{aligned}
p(c_1,a_1) &= 0.6, &\quad p(c_1,a_2) &= 0.1,\\[2pt]
p(c_2,a_1) &= 0.1, &\quad p(c_2,a_2) &= 0.2.
\end{aligned}
\]
\paragraph{Marginal probabilities.}
Summing over the complementary subsystem yields
$
p(c_1)  = p(a_1) = 0.7, p(c_2) = p(a_2)=0.3.
$

\paragraph{Conditional probabilities.}
Using \(p(c_i\!\mid\!a_j)=p(c_i,a_j)/p(a_j)\) and \(p(a_j\!\mid\!c_i)=p(c_i,a_j)/p(c_i)\), we obtain

\noindent\emph{Conditionals of cognition given affect:}
$$
p(c_1\mid a_1) = \frac{6}{7} \approx 0.8571,
p(c_2\mid a_1)  \frac{1}{7} \approx 0.1429, p(c_1\mid a_2) = \frac{1}{3} \approx 0.3333,
p(c_2\mid a_2) = \frac{2}{3} \approx 0.6667.
$$

\noindent\emph{Conditionals of affect given cognition:}
$$
p(a_1\mid c_1) = \frac{6}{7} \approx 0.8571,
p(a_2\mid c_1)= \frac{1}{7} \approx 0.1429,
p(a_1\mid c_2)= \frac{1}{3} \approx 0.3333,
p(a_2\mid c_2) = \frac{2}{3} \approx 0.6667.
$$

\paragraph{Interpretation.}
Unlike the perfect-correlation  example (where matching pairs had conditional probability \(1\) and cross pairs \(0\)), here the conditional probabilities are between \(0\) and \(1\). For instance, observing affect \(a_1\) makes cognitive \(c_1\) much more likely than \(c_2\) (\(p(c_1\mid a_1)\approx0.8571\)), but it does not determine cognition uniquely. Similarly, observing cognition \(c_2\) makes affect \(a_2\) more likely (\(p(a_2\mid c_2)\approx0.6667\)) without full certainty. This toy state therefore models \emph{partial} cognitive–affective coupling (partial entanglement) suitable for illustrating graded coherence in SLT.

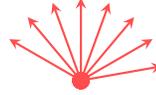
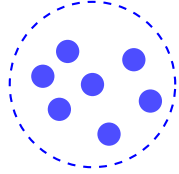
\begin{figure}[h!]
\centering
\begin{tikzpicture}[scale=1.1]

\node at (0,2.2) {\textbf{Social Cohesion}};
\foreach \x/\y in {0/0,0.5/0.3,-0.3/0.4,0.7/-0.2,-0.4/-0.3,0.2/-0.6,-0.6/0.1} {
  \fill[blue!70] (\x,\y) circle (4pt);
}

\draw[blue, thick, dashed] (0,0) circle (1);

\node at (6,2.2) {\textbf{Social Coherence}};
\foreach \angle in {10,30,50,70,90,110,130,150} {
  \draw[red!70, thick, -stealth] (6,0) -- ++(\angle:1);
  \fill[red!70] (6,0) circle (3pt);
}

\end{tikzpicture}
\caption{Illustration of the difference between social cohesion (left) and social coherence (right).
On the left, cohesion is shown as dots clustered together (bound by a dashed circle). On the right, coherence is shown as arrows pointing in aligned directions from a common center (like phase-synchronized emission in a laser).}

\end{figure}

\section*{Acknowledgments} We would like to thank  A. Alodjants and M. Shutova for fruitful discussions which played teh important role 
in development of the present model coupling SLT and theory of mental markers.   
The  research of one of the authors (A.Kh.) was partially supported by  the EU-grant CA21169 (DYNALIFE)  and by the JST and CREST Grant Number JPMJCR23P4, Japan.

\end{document}